\documentstyle [11pt] {article}

\renewcommand{\theequation}{\thesection.\arabic{equation}}
\newcommand{\sect}[1]{ \section{#1} \setcounter{equation}{0} }

\newcommand{\req}[1]{(\ref{#1})}

\newcommand{\m}{\hat{p}}
\newcommand{\Z}{\ZZ}
\newcommand{\nwc}{\newcommand}
\nwc{\kae} {K\"{a}hler }
\nwc{\hyp} {\hyphenation}
\nwc{\be}  {\begin{equation}}
\nwc{\ee}  {\end{equation}}
\nwc{\ba}  {\begin{array}}
\nwc{\ea}  {\end{array}}
\nwc{\bdm} {\begin{displaymath}}
\nwc{\edm} {\end{displaymath}}

\nwc{\bea} {\be\ba{lcl}}
\nwc{\eea} {\ea\ee}

\nwc{\bda} {\bdm\ba{lcl}}
\nwc{\eda} {\ea\edm}

\nwc{\bc}  {\begin{center}}
\nwc{\ec}  {\end{center}}

\nwc{\ds}  {\displaystyle}
\nwc{\bmat}{\left(\ba}
\nwc{\emat}{\ea\right)}
\nwc{\nn} {\nonumber}
\nwc{\nnn} {\nonumber \vspace{.2cm} \\ }
\nwc{\ra}{\rightarrow}
\nwc{\lra}{\longrightarrow}

\nwc{\p} {\partial}
\nwc{\SS} {S}

\nwc{\sieb}{\bf \overline{27}}

\nwc{\scr}  {\scriptstyle}
\nwc{\tx}  {\textstyle}
\nwc{\scs} {\scriptscriptstyle}

\nwc{\ov}  {\overline}

\nwc{\hb}  {\bar h}
\nwc{\xb}  {\bar x}
\nwc{\yb}  {\bar y}
\nwc{\zb}  {\bar z}
\nwc{\wb}  {\bar w}
\nwc{\Ob}  {\bar O}
\nwc{\Yb}  {\bar Y}

\nwc{\ep} {\epsilon}
\nwc{\de} {\delta}
\nwc{\Th} {\Theta}
\nwc{\th} {\theta}
\nwc{\al} {\alpha}
\nwc{\si} {\sigma}
\nwc{\om} {\omega}
\nwc{\Om} {\Omega}
\nwc{\Ga} {\Gamma}
\nwc{\ga} {\gamma}
\nwc{\bet} {\beta}

\nwc{\Sc}  {{\cal S}}
\nwc{\Rc}  {{\cal R}}
\nwc{\Dc}  {{\cal D}}
\nwc{\Oc}  {{\cal O}}
\nwc{\Cc}  {{\cal C}}
\nwc{\gc}  {{\cal g}}

\nwc{\Of}  {{\cal O}_f}
\nwc{\Oft} {{\cal O}_{f_2}}
\nwc{\Ofo} {{\cal O}_{f_1}}

\nwc{\Pc}  {{\cal P}}
\nwc{\Mc}  {{\cal M}}
\nwc{\Ec}  {{\cal E}}
\nwc{\Fc}  {{\cal F}}
\nwc{\Hc}  {{\cal H}}
\nwc{\Kc}  {{\cal K}}
\nwc{\Wc}  {{\cal W}}

\nwc{\Fcp} {{\cal F}^\pr}
\nwc{\Hcp} {{\cal H}^\pr}

\nwc{\Xc}  {{\cal X}}
\nwc{\Gc}  {{\cal G}}
\nwc{\Zc}  {{\cal Z}}
\nwc{\Nc}  {{\cal N}}

\nwc{\xc}  {{\cal x}}
\nwc{\Ac}  {{\cal A}}
\nwc{\Bc}  {{\cal B}}
\nwc{\Uc} {{\cal U}}
\nwc{\Vc} {{\cal V}}
\nwc{\Lc} {{\cal L}}
\nwc{\Qc} {{\cal Q}}

\nwc{\lng} {\langle}
\nwc{\rng} {\rangle}

\nwc{\lf} {\left}
\nwc{\ri} {\right}

\nwc{\pr} {\prime}
\nwc{\diag} {{\rm diag}}
\nwc{\inv}  {{\rm inv}}
\nwc{\mod}  {{\rm mod}}
\nwc{\dete}  {{\rm det}}
\nwc{\tr}  {{\rm tr}}
\nwc{\im}  {{\rm Im}}
\nwc{\re}  {{\rm Re}}

\nwc{\h} {\frac{1}{2}}
\nwc{\fc} {\frac}

\def\KK{\relax{\rm I\kern-.18em K}}
\def\RR{\relax{\rm I\kern-.18em R}}
\def\NN{\relax{\rm I\kern-.18em N}}
\def\PP{\relax{\rm I\kern-.18em P}}

\def\zz{\relax{\sf Z\kern-.3em Z}}
\def\ZZ{\relax{\sf Z\kern-.4em Z}}
\def\ZZZ{Z\kern -0.31em Z}
\def\QQ{{\rm \kern .25em
             \vrule height1.4ex depth-.12ex width.06em\kern-.31em Q}}
\def\CC{{\rm \kern .25em
             \vrule height1.4ex depth-.12ex width.06em\kern-.31em C}}
\if@twoside\oddsidemargin25mm
\else\oddsidemargin25mm\evensidemargin25mm\marginparwidth25mm\fi%
\begin{document}

\begin{titlepage}
\begin{flushright}MPI--Ph/93--07 \\ TUM--TH--152/93 \\ hep-th/9303017 \\ \ \\
February 1993
\end{flushright}
\vfill

\begin{center}
{\Large \bf Threshold Corrections to Gauge Couplings \\
in Orbifold Compactifications$^\ast$}   \\
\vskip 1.2cm
{\large \bf P. Mayr}\ \ and\ \ {\large \bf S. Stieberger} \\
\vskip .5cm

{\large \em Physik Department} \\
{\large \em Institut f\"{u}r Theoretische Physik} \\
{\large \em Technische Universit\"at M\"unchen} \\
{\large \em D--8046 Garching, FRG}
\vskip .25cm
{\rm and}\\
\vskip .25cm
{\large \em Max--Planck--Institut f\"ur Physik} \\
{\large \em ---Werner--Heisenberg--Institut---}\\
{\large \em P.O. Box 401212}\\
{\large \em D--8000 M\"{u}nchen, FRG}
\end{center}
\vfill
\vspace{4cm}

\begin{center}
{\bf ABSTRACT}
\end{center}

\begin{quote}
We derive the moduli dependent threshold corrections to gauge couplings in
toroidal orbifold
compactifications. The underlying six dimensional torus lattice of the
heterotic string theory is not assumed
---as in previous calculations--- to decompose
into a direct sum of a four--dimensional and a two--dimensional sublattice,
with
the latter lying in a plane left fixed by a set of orbifold twists.
In this more general case the threshold corrections are no longer
automorphic functions of the modular group,
but of certain congruence subgroups of the modular group. These groups can also
be obtained by studying the massless spectrum; moreover they have larger
classes
of automorphic functions. As a consequence the threshold corrections cannot
be uniquely determined by symmetry considerations and
certain boundary conditions at special points in the moduli space,
as was claimed in previous publications.

\vskip 5mm \vskip0.5cm
\hrule width 5.cm \vskip 1.mm
{\small\small
\hspace{-1.1cm} \noindent $^\ast$ Supported by the Deutsche
Forschungsgemeinschaft
and the EC under contract SC1--CT92--0789.}
\normalsize
\end{quote}


\end{titlepage}

\sect{Introduction}

String theory is the only known theory which consistently unifies all
interactions. To make contact with the observable world
one constructs the field--theoretical low--energy limit
of a given ten--dimensional string theory.  To get a
four--dimensional effective N=1 supergravity theory one compactifies six
of the ten dimensions on an internal
Calabi--Yau manifold \cite{chsw} or a toroidal orbifold \cite{dhvw}.
In addition one integrates out
all the massive string modes corresponding to excited string states as well as
states with momentum
or winding numbers in the internal dimensions. The resulting theory
then describes the
physics of the massless string excitations at low energies in
field--theoretical
terms. One question of current interest are threshold corrections to the
gauge couplings due to the infinite tower of massive string modes.
Since they provide the boundary conditions for the running gauge couplings
they are the foundation of any discussion about gauge coupling unification.

One property of string theory is that there are infinitely many vacua
labeled by the vacuum expectation value of certain scalar fields:
the dilaton field and the moduli fields. The latter parametrize
the shape and the size of the compactification manifold. They enter
the low--energy theory as massless scalar fields with flat effective
potentials neglecting non--perturbative effects. The moduli space of
string vacua is invariant under certain discrete reparametrizations,
e.g. target space duality.
It is clear that the limiting effective supergravity theory should reflect
these stringy symmetries. In particular it has to be anomaly free
with respect to the discrete reparametrizations.
Since the supergravity Lagrangian for the massless fields
produces anomalous triangle
graphs \cite{dfkz,l,co}, anomaly freedom has to be restored by adding
appropriate
counterterms. Indeed the full string theory generates two kinds
of such counterterms: an universal gauge--group
independent Green--Schwarz term and the threshold corrections
arising from the contribution of heavy string modes in the loops \cite{dfkz}.

The moduli dependence of the effective action can be
determined by calculating the
appropriate string scattering amplitudes.
In particular this point has been discussed for the
\kae potentials \cite{DKL1},
thresholds corrections to the gauge couplings \cite{kap} as well as gravitional
and Yukawa
couplings \cite{anton,agn} in a general framework.
More concrete results can be obtained for orbifold
compactifications: This issue was discussed for the
\kae potentials \cite{clo}, Yukawa couplings \cite{ejss} and
threshold corrections
to the gauge couplings due to the influence of the
massive string modes \cite{DKL2,agn}.
Threshold corrections to the gauge couplings play a crucial r\^{o}le for
anomaly cancellation in the effective Lagrangian \cite{dfkz} as well as
for string unification \cite{ILR,IL}.
The modular invariant threshold corrections to the gauge couplings
evaluated in ref. \cite{DKL2} can be considered to split into two
parts: The first is interpreted to be that of the massless states and
is the analogue of
the field theoretical triangle graph involving two gauge fields and one
auxiliary connection corresponding to either the \kae connection or the
sigma model connection \cite{dfkz,l,co}. This term transforms anomalous
under the discrete reparametrizations on the moduli space. The
other part is holomorphic and describes a modification of the effective
Lagrangian due to the
contribution of
super--heavy string modes in the
loops. From the point of field theory this term shows up
as a counterterm repairing the duality anomaly of the first term.
On the other hand the threshold corrections provide the boundary
conditions for the renormalization group equations of the
running  gauge
couplings and from the field--theoretical point of view thus determine
the unification mass $M_{\rm X}$. They may shift $M_{\rm X}$
to the string scale $M_{\rm string}$ or even to higher values
\cite{ILR,IL}. Finally the threshold corrections to the gauge
couplings may also be important for supersymmetry breaking via gaugino
condensation \cite{susybr}.

The effective field theory has in general singularities because
one has also integrated over
massive states which become massless at special points in the
moduli space. The large radius behaviour $\im \
T\rightarrow \infty$ of the moduli
dependence
of the one--loop corrections to the gauge couplings was first discussed in
ref. \cite{in}. In this limit an infinite number of Kaluza--Klein states
becomes massless and causes a singularity in the threshold
corrections linear in $\im \ T$. On the other hand the winding
states become massless at the duality transformed point $\im \ T=0$.
There can be also special points in the moduli space
where additional charged fields become massless and give rise to logarithmic
poles in the threshold function. The most general function which reproduces
this singular behaviour and is in addition target--space modular
invariant is given  by the expression \cite{cfilq}:

$$\triangle(T,\bar T)= -b \ln\lf[\im T |\eta(T)|^4\ri]
+\fc{b}{3} \ln |H(T)|^2\ ,$$
where $\eta$ is the Dedekind eta-function and  $b$ is related to
the one--loop $\bet$--function in a way to be specified later.
$H(T)$ is a modular invariant function which has no zeros and poles at
$\im \ T\rightarrow \infty$ and determines the functional behaviour at the
critical
points.

In refs. \cite{ov,fklz} it was shown that the functional dependence of the
threshold corrections to the gauge couplings is the same as the one of the
topological free energy of that two--dimensional subspace of the full torus
lattice, which is left fixed by the orbifold twist. This topological
invariant is believed
to be invariant under the modular group $\Ga \equiv SL(2,\ZZ)$ and
thus should be
an automorphic function of $\Ga$.

An explicit expression for the moduli dependent threshold correction to the
gauge couplings in the case of orbifold
compactifications was derived in \cite{DKL2}. Here the sum
over all massive string modes was evaluated.
It was shown there that only orbifold sectors with one unrotated plane
contribute to this function. Since
these sectors posses two space--time supersymmetries they are called
N=2 sectors.
In ref. \cite{DKL2,agn} only the special case was considered,
where the six--dimensional lattice splits into a direct sum of a
two--dimensional
and a four--dimensional lattice $T^6 = T^2 \oplus T^4$ with the unrotated
plane lying in $T^2$. The main purpose of this paper is to generalize these
calculations
to the larger class of orbifold models which do not fulfill this
assumption.
In fact most of the toroidal orbifolds are of this
type \cite{ek}.

The new feature is the fact that the
lattice  vectors characterizing momentum and
winding states can now have non--vanishing components in all
lattice directions. This has to be contrasted with the investigations of refs.
\cite{DKL2,fklz} where these states had only two non--vanishing components.
This novelty has an important consequence: The threshold corrections are no
longer invariant under the full modular group $\Ga$ but only under a certain
subgroup of $\Ga$. Since those subgroups have in general
a larger set of automorphic functions, the thresholds will be composed of
different terms. As required by consistency, also the topological free
energy for these kind of orbifolds shares this property and therefore
is no longer
invariant under the full modular group.

The organization of this paper is as follows: In section 2
we present the basic formulas necessary to determine the threshold
corrections for the general case of orbifolds with no restrictions
on the lattice. Then we evaluate the
threshold corrections to the gauge couplings for a $Z_4,\ Z_6$ and $Z_8$
orbifold. We compare the results with the corresponding topological free
energies and determine the massless spectrum.
As a consequence of the restriction of the threshold symmetry
to a subgroup of $\Ga$ the duality anomaly cancellation
terms in the effective action are modified, too.
Therefore the boundary conditions for the running
gauge couplings and the unification mass change. This will be discussed
in section 3.

\sect{Threshold corrections to the gauge couplings for orbifold string vacua}

The starting point for calculating the one--loop threshold corrections
$\triangle_a$ to the inverse gauge coupling $g_a^{-2}$ for orbifold
compactifications is the general formula of
\cite{kap}, valid for any four--dimensional tachyon--free vacuum of the
heterotic string

\be
\triangle_a= \int_{\Fc} \fc{d^2 \tau}{\tau_2} [\Bc_a(\tau,\bar \tau)-b_a]\ ,
\label{threth}
\ee
where
\bea
\Bc_a(\tau,\bar \tau) &=& \ds{2 |\eta(\tau) |^{-4} \sum_{{\rm even} \ s}
(-1)^{s_1+s_2}
\fc{{\rm d} Z_{\psi}({\bf s},\bar \tau)}{2 \pi i {\rm d} \bar \tau}
\tr_{s_1}\lf[Q_a^2 (-1)^{s_2 F} q^{H-\fc{11}{12}} \bar q^{\bar H
-\fc{3}{8}}\ri]_{\rm int} \ ,}\nnn
b_a&=& \ds{\lim_{\tau_2 \ra \infty} \Bc_a(\tau,\bar \tau)=-\fc{11}{3} \tr_{\rm
V}Q_a^2+\fc{2}{3}\tr_{\rm F}Q_a^2+\fc{1}{3}\tr_{\rm S}Q_a^2\ .}
\label{thre}
\eea
Here $\tau =\tau_1+i \tau_2$ is the modulus of the world--sheet torus and
 $\Fc$ is the
fundamental region $\Fc=\{\tau \in \CC\ |\ \tau_2 >0 , |\tau_1| < \h,
|\tau|>1 \}\ $. Furthermore $ q=e^{2 \pi i \tau} $ and $Q_a$ is the
charge of a state with respect to a generator of the gauge group labeled by
$a$.
The spin
structure of the fermions is denoted by  ${\bf
s}=(s_1,s_2)$, where $s_1,s_2 \in \{0,1\}$. A zero (one) refers to
anti-periodic (periodic) boundary condition on the torus.
The partition function
for one complex fermion is given by

\be
Z_{\psi}({\bf s},\bar \tau)= \fc{1}{\eta(\bar \tau)} \th \lf[s_2/2 \atop
s_1/2\ri]\ .
\ee
The sum over the spin structures excludes the parity odd contribution ${\bf
s}=(1,1)$. Finally $b_a$ is related to the
one--loop $\bet$--function via $\beta_a = b_ag^3_a/16\pi^2$.

In ref. \cite{DKL2} the dependence of $\triangle_a$ on the internal moduli has
been evaluated for certain orbifold compactifications.
Such vacua are obtained \cite{dhvw} by moding the six--dimensional torus
$T^6=\RR^6/\Lambda$ of the compactification manifold
by a twist $\Th$
of finite order ($\Th^N=1$ for some $N$); in addition $\Th$ has to be an
automorphism of the lattice $\Lambda$.
In refs. \cite{kap,DKL2} it was argued that
the moduli dependence of the
threshold corrections \req{thre} arises only from the orbifold sectors
with N=2 space--time supersymmetry.
Therefore we will not consider the completely untwisted sector
with N=4 space--time supersymmetry as well as N=1 supersymmetry
preserving sectors.
The orbit of these N=2 sectors will be denoted by $\Oc$ and is
represented by the relevant set of
twisted boundary conditions along
the two cycles of the world--sheet torus.
One property of N=2 sectors is that they
always have a complex plane left fixed by the orbifold
twist $\Th$. For calculational reasons it was assumed in ref. \cite{DKL2}
that these fixed planes lie in a two dimensional
even and self--dual sublattice $\Lambda_2$ of the full torus
lattice $\Lambda$. Therefore their results can be applied only to
models where
the torus lattices $\Lambda$
can be written as an orthogonal decomposition of $\Lambda$
into  a four dimensional lattice $\Lambda_4 $ and a two dimensional
lattice $\Lambda_2$, with the fixed plane lying in $\Lambda_2$.
On the other hand there are a lot of
orbifold models which do not fulfill this condition \cite{ek};
in the following we
present the calculation of the threshold corrections
for this larger class of orbifold models.

In ref. \cite{DKL2} only the difference
$\triangle_{a_1}/k_{a_1}-\triangle_{a_2}/k_{a_2}$ between two gauge groups was
calculated to get rid
off the non--modular invariant regulator terms. In particular the
difference $(\tau_2/k_{a_1})\Bc_{a_1}-(\tau_2/k_{a_2})\Bc_{a_2}$ is
invariant under transformations of the modular group
$\Ga$ acting on $\tau$ with generators $S: \tau \ra -1/\tau$
and $T: \tau \ra \tau+1$. Moreover it is easy to realize that the N=2
part of the above difference is separately invariant under $\Ga$:
this is because the generators $S$ and $T$ change the boundary conditions
$(\Th^m,\Th^n)$ of a orbifold sector
by $(\Th^m,\Th^n) \ra (\Th^n,\Th^{N-m})$ and $(\Th^m,\Th^n)  \ra
(\Th^m,\Th^{m+n})$, respectively.
The N=2 orbit $\Oc$ can be created from
so--called fundamental elements by acting on them with
modular transformations. The
set of fundamental elements
will be denoted by $\Oc_0$. Instead of integrating
$(\tau_2/k_{a_1})\Bc_{a_1}-(\tau_2/k_{a_2})\Bc_{a_2}$
over the integration region $\Fc$ as in \req{threth}, the sum
over all N=2 sectors
$\Oc$ can be replaced by the sum over the fundamental elements $\Oc_0$,
provided that the integration region is
extended to a region $\tilde \Fc$. Here $\tilde \Fc$ is generated by
acting on $\Fc$ with exactly those modular transformations
which generate $\Oc$ from $\Oc_0$. The integrand is then invariant only
under a congruence subgroup $\Ga_0$ of $\Ga$ \cite{apo,moore} and $\tilde \Fc$
is its fundamental region.
As an example consider the $Z_4$ orbifold generated by a twist
with complex eigenvalues $\Th=(i,i,-1)$. The three N=2 sectors are
$\Fc=\{(1,\Th^2),(\Th^2,1),(\Th^2,\Th^2)\}$. All these sectors can be generated
from the fundamental element $(1,\Th^2)$ by an S-- and an ST--transformations,
 respectively.
The appropriate extension of the integration region is then given by
$\tilde \Fc=\{1,S,ST\}\Fc$.

In general there is no known way of calculating integrals over the
fundamental region of modular functions analytically.
Fortunately the $\tau$--integration over
$(\tau_2/k_{a_1})\Bc_{0a_1}-(\tau_2/k_{a_2})\Bc_{0a_2}$\footnote{The zero of
$\Bc_0$
means that the trace in \req{thre} is restricted to $\Oc_0$.}
simplifies to an integration only over the zero mode part
$Z^{torus}_{(g,h)}(\tau,\bar \tau)$ corresponding to the
winding and Kaluza--Klein modes of the fixed torus. Here
$(g,h)$ denotes the dependence of this function on the orbifold
sector; we will
give an explicit expression below. This result can
be seen in two different ways:
first one can consider the explicit representation of the integrand in terms of
$\th$--functions.
The charge operators in the trace over the internal part
can be represented by $\tau$--derivatives on the appropriate
$\th$--functions of the gauge sector \cite{ellis}.
Indeed the derivative of $Z_\psi$ in \req{thre} has a similar
source. In this way the integrand splits into two parts:
first $Z^{torus}_{(g,h)}(\tau,\bar \tau)$
representing the zero--modes of that
part of the lattice which is left unrotated by the twist.
This part contains the full moduli dependence and multiplied by
$\tau_2$ it is invariant under a
subgroup $\Ga_0$ of $\Ga$. The other part describes  the
rest of the right--moving and the left--moving sector of a string
twisted by $(g,h)$. It is separately invariant
under $\Ga_0$ and bounded in the fundamental region $\tilde \Fc$ of
$\Ga_0$. For the relevant subgroups
the only automorphic functions without singularities in the whole
fundamental region are constants due to general theorems \cite{apo}.
What will remain is therefore an integral over the
zero--mode parts $Z^{torus}_{(g_0,h_0)}(\tau,\bar \tau)$, where $(g_0,h_0)
\in \Oc_0$, with integration
region  $\tilde \Fc$ multiplied by a constant expression to be
determined.

Actually there is a second more elegant way to derive this result,
namely a generalization of the method of ref. \cite{DKL2}.
The main difference is that
$Z^{torus}_{(g,h)}(\tau,\bar \tau)$
depends now on the orbifold sector $(g,h)$, in contrast to the
orbifold models considered in \cite{DKL2}. The N=2 orbit of interest
can be interpreted as an N=2
supersymmetric model in four dimensions. This is because this orbit
itself would produce an orbifold
with N=2 space--time supersymmetry.
The internal part of a theory with an N=2 spacetime supersymmetry splits
into one piece with an N=2 and a second piece with an N=4
superconformal symmetry \cite{bd}. Moreover it
couples to the spin structure of the world--sheet fermions only via
an $SU(2)$ current \cite{bd}. This fact
can be used to rewrite $\Bc_a(\tau,\bar \tau)$ of \req{thre} as follows:

\bea
\Bc_a(\tau,\bar \tau) &=& \ds{\sum_{(g,h) \in \Oc}Z^{torus}_{(g,h)}(\tau,\bar
\tau)\
\Cc^{(g,h)}_a(\tau)\ ,}\nnn
\Cc^{(g,h)}_a(\tau)&=& \ds{\eta^{-4}(\tau) \tr_R\lf[\h (-1)^{\rm F} Q_a^2
q^{H-\fc{5}{6}} \bar q^{\bar H-\fc{1}{4}}\ri]_{(g,h) \atop (c,\bar c)=(20,6)}\
.}
\label{nata}
\eea

Here $\Cc^{(g,h)}_a(\tau)$
contains the $E_8 \times E_8$ current algebra twisted by $(g,h)$,
the contributions of the four internal twisted dimensions,
the oscillator part of the fixed torus and the contribution of the
four space--time dimensions.

Now $\tau_2Z_{(g_0,h_0)}^{torus}(\tau,\bar \tau)$ is invariant under
$\Ga_0$--transformations on $\tau$.Since
$(\tau_2/k_{a_1})\Bc_{a_1}-(\tau_2/k_{a_2})\Bc_{a_2}$ is
invariant under $\Ga$ and therefore in particular under $\Ga_0$,
we can follow that the function
$1/k_{a_1}\Cc^{(g_0,h_0)}_{a_1}(\tau)-1/k_{a_2}\Cc^{(g_0,h_0)}_{a_2}(\tau)$
has to be invariant under $\Ga_0$ as well. Moreover it is nowhere singular
as can be seen from the limit $\tau_2 \rightarrow \infty$.
Therefore it has to be a constant by the virtue of the above mentioned
theorems.
This constant represents the contribution $b_a^{(g,h)}$ of the
sector $(g,h)$ to the one--loop $\bet$--function coefficient
$b_a$. Including an infrared regularization \cite{kap},
the threshold corrections
\req{thre} can be expressed by the following integral:

\bea
\triangle_a &=& \ds{\int \limits_{\Fc}
\fc{d^2\tau}{\tau_2}\lf[\sum_{(g,h) \in \Oc}\fc{Z_{(g,h)}^{torus}(\tau,\bar
\tau)}{\lim \limits_{\tau'_2 \rightarrow \infty}
Z_{(g,h)}^{torus}(\tau',\bar\tau')}
b_a^{(g,h)}-b_a^{N=2}\ri]}\nnn
&=&\ds{\sum_{(g_0,h_0) \in
\Oc_0} b_a^{(g_0,h_0)} \int \limits_{\tilde \Fc}
\fc{d^2\tau}{\tau_2}
Z_{(g_0,h_0)}^{torus}(\tau,\bar
\tau)-b_a^{N=2}\int
\limits_\Fc \fc{d^2\tau}{\tau_2}\ .}
\label{thresh}
\eea
Here $b_a^{N=2}$ denotes the $\bet$--function coefficient
arising from the N=2 sectors of
the $Z_N$ orbifold under consideration and we have used the identity $\lim
\limits_{\tau_2 \rightarrow \infty} Z_{(g_0,h_0)}^{torus}(\tau,\bar\tau')=1$.
The first integral on the r.h.s
represents a
Petersson product on the space of modular functions of the corresponding
subgroup \cite{moore,leh}.

\subsection{Threshold corrections for the $SU(4)^2/Z_4$ orbifold}
\label{z4}

In the rest of this section we will apply the previous
analysis to some orbifold models which can not be treated
with the method of ref. \cite{DKL2}.
First we consider the $Z_4$ orbifold defined by two Coxeter twists
in two root lattices of
$SU(4)$ and deformations of them. The twist $Q$ with the complex eigenvalues
$\Th=(i,i,-1)$ can be
represented w.r.t to the lattice basis ($e'_i=Q_i^{\ j}\ e_j$) by the matrix:

\be
Q=\left(\ba{ccccccr}
0&0&-1&0&0&0 \\
1&0&-1&0&0&0 \\
0&1&-1&0&0&0 \\
0&0&0&0&0&-1 \\
0&0&0&1&0&-1 \\
0&0&0&0&1&-1
\ea \right)\ .
\ee
There are six complex untwisted moduli: five $(1,1)$--moduli and one
$(2,1)$-modulus due to the five untwisted $\sieb$ generations and one untwisted
$\bf 27$ generation \cite{dhvw,ek}. Therefore the metric $g$ (defined by
$g_{ij}=<e_i|e_j>$) has seven and the antisymmetric
tensor field $b$ five real deformations. The equations
$gQ=Q^{\ast}g$ and $bQ=Q^{\ast}b$ determine the background
fields in terms of the independent deformation parameters\footnote{We do not
consider the more
general case with a non--commuting $b$ field \cite{MS2}.} \cite{MS1,MS2}.
Solving
these equations one obtains

\be
g=\left(\ba{ccccccr}
R^2&x&-R^2-2x&u&v&-u-v-z\\
x&R^2&x&z&u&v\\
-R^2-2x&x&R^2&-u-v-z&z&u\\
u&z&-u-v-z&S^2&y&-S^2-2y\\
v&u&z&y&S^2&y\\
-u-v-z&v&u&-S^2-2y&y&S^2
\ea \right)\ ,
\ee
with
$R,S,x,y,u,v,z \in \RR$ and

\be
b=\left(\ba{ccccccr}
0&\al&0&\ga&\de&-\ga-\de-\ep\\
-\al&0&\al&\ep&\ga&\de\\
0&-\al&0&-\ga-\de-\ep&\ep&\ga\\
-\ga&-\ep&\ga+\de+\ep&0&\bet&0\\
-\de&-\ga&-\ep&-\bet&0&\bet\\
\ga+\de+\ep&-\de&-\ga&0&-\bet&0
\ea \right)\ ,
\ee
with $\al,\bet,\ga,\de,\ep \in \RR\ .$
%
%
%
%

As mentioned previously the N=2 orbit is given by $\Oc=\{(1,\Th^2),(\Th^2,1),
(\Th^2,\Th^2)\}$.
$(\Th^2,1)$ can be obtained from the {\em fundamental element} $(1,\Th^2)$
by an $S$--transformation on $\tau$ and
similarly $(\Th^2,\Th^2)$ by an $ST$--transformation.
The zero mode parts $Z_{(g,h)}^{torus}$ of the fixed plane take
the following form \cite{ek}:

\bea
Z_{(1,\Th^2)}^{torus}(\tau,\bar \tau,g,b) &=& \ds{\sum\limits_{P \in
\Lambda_N^\bot}
      q^{\h P_L^2} \bar q^{\h P_R^2}\ ,}   \nnn
Z_{(\Th^2,1)}^{torus}(\tau,\bar
\tau,g,b)&=&\ds{\fc{1}{V_{\Lambda_N^\bot}}\sum\limits_{P \in
(\Lambda_N^\bot)^\ast}
   q^{\h P_L^2} \bar{q}^{\h P_R^2} \ ,}\nnn
Z_{(\Th^2,\Th^2)}^{torus}(\tau,\bar
\tau,g,b)&=&\ds{\fc{1}{V_{\Lambda_N^\bot}}\sum\limits_{P
\in(\Lambda_N^\bot)^\ast}
   e^{\pi i (P_L^2 - P_R^2)} \; q^{\h P_L^2} \bar{q}^{\h P_R^2}\ .}
\label{zero}
\eea
Here $\Lambda_N$ denotes the Narain lattice of $SU(4)^2$ with
momentum vectors \cite{NSW}

\be
P_L = \ds{\fc{p}{2}+(g-b)w\ ,} \hspace{1cm}
P_R = \ds{\fc{p}{2}-(g+b)w\ .}
\ee
$\Lambda_N^\bot$ is that part of the lattice which remains fixed under
$Q^2$ and $V_{\Lambda^\bot_N}$ is the volume of this sublattice.
It is worth mentioning that the partition functions
$Z_{(g,h)}^{torus}(\tau,\bar \tau,g,b)$ of ref. \cite{DKL2}
are independent of the orbifold sector $(g,h)$ because there
$\Lambda^\bot_N$ is an even self--dual sublattice. As a
consequence $\tau_2Z_{(g,h)}^{torus}(\tau,\bar \tau,g,b)$ was
invariant under $\Ga$ in contrast to the functions of eq.
(\ref{zero}).

The subspace corresponding to $\Lambda^\bot_N$
can be described by the following winding and momentum vectors,
respectively:

\be
w=\lf(\ba{c} n^1\\0\\n^1\\n^2\\0\\n^2
\ea \ri)\ ,\ \ n^1,n^2 \in \ZZ\ \ \ {\rm and}\ \ \ p=\lf(\ba{c}
m_1\\-m_1\\m_1\\m_2\\-m_2\\m_2\ea \ri)\ ,\ \ m_1,m_2 \in \ZZ\ .
\label{vecz4}
\ee
They are determined by the equations $Q^2w=w$ and
$Q^{\ast^2}p=p$.
The partition function
$\tau_2Z_{(1,\Th^2)}^{torus}(\tau,\bar \tau,g,b)$ is invariant under the group
$\Ga_0(2)$
which belongs to the congruence subgroups of $\Ga$ \cite{apo}.
$\Ga_0(2)$ is generated by the elements $T$ and $ST^2S$ of $\Ga$.
In general the group $\Ga_0(n)$ can be
represented by the following set of two by two matrices
acting on $\tau$ as $\tau \ra \fc{a\tau+b}{c\tau+d}$:
\be
\Ga_0(n)=\lf\{\lf( \ba{lr}  a&b\\ c&d \ea \ri)\ |\ ad-bc=1, c=0\ \mod\ n\ri\} \
{}.
\label{group}
\ee

Instead of integrating the contribution of the various
sectors $(g,h)$ over the fundamental region $\Fc$, we integrate
only the contribution $Z_{(1,\Th^2)}^{torus}$ but over the
extended integration region $\Fc=\{1,S,ST\}\Fc$ (cf. \req{thresh}).
The explicit expression for $Z_{(1,\Th^2)}^{torus}$ can
be determined
with \req{vecz4} to be

\bea
\ds{Z_{(1,\Th^2)}^{torus}(\tau,\bar \tau,g,b)} &=& \ds{\sum_{(P_L,P_R) \in
\Lambda_N^\bot} q^{\h P_L^tg^{-1}P_L}\ \bar
q^{\h P_R^tg^{-1}P_R}} \nnn
&=& \ds{\sum_{p,w}e^{2 \pi i \tau p^t w} e^{-\pi\tau_2
(\h p^tg^{-1}p-2p^tg^{-1}bw+2w^tgw-2w^tbg^{-1}bw-2 p^tw)}\ .}
\label{general}
\eea
The last expression can be written in terms of the (1,1)--modulus $T$ and the
(2,1)--modulus $U$ corresponding to $\p Y^3(z)
$ and $ \bar \p
Y^3(\bar z)$ \cite{dvv}. They are defined by:

\be
T=T_1+iT_2=2(b+i \sqrt{\dete g_\bot})\ \ ,\ \
U=U_1+iU_2=\fc{1}{g_{\bot 11}}(g_{\bot 12}+\sqrt{\dete g_\bot})\ .
\ee
Here $g_\bot$ is uniquely determined by $w^tgw=(n^1 n^2)g_\bot\lf(n^1 \atop n^2
\ri)$. This way one gets

\bea
T &=& \ds{4 \lf[ -\de-\ep+i \sqrt{4 xy-(v+z)^2} \ri]\ ,} \nnn
U &=& \ds{\fc{1}{2x} \lf[v+z-i \sqrt{4 xy-(v+z)^2} \ri]\ .}
\label{complexmod}
\eea
This definition of $T$ and $U$ agrees with that appearing in
the \kae potential \cite{clo,DKL1}. The partition function
$Z_{(1,\Th^2)}^{torus}(\tau,\bar \tau,g,b)$ takes now the form

\be
Z_{(1,\Th^2)}^{torus}(\tau,\bar \tau,T,U)= \sum_{m_1,m_2 \in 2 \ZZZ \atop
n^1,n^2 \in \ZZZ}e^{2 \pi i \tau (m_1
n^1+m_2n^2)} e^{\fc{-\pi \tau_2}{T_2U_2} |T U n^2+T
n^1-Um_1+m_2|^2}\ .
\label{partition}
\ee
This expression can be rewritten performing
a Poisson resummation on $m_1$ and $m_2$:

\be \label{syz4}
\ba{lrlcr}
&\tau_2 \ Z_{(1,\Th^2)}^{torus}(\tau,\bar \tau,T,U)&=& \ds{\fc{1}{4} \sum_{A
\in
\Mc} e^{- 2 \pi i T \dete A}\ T_2\
e^{\fc{-\pi T_2}{\tau_2 U_2} \lf|(1,U)A \lf(\tau \atop 1\ri)\ri|^2}\ ,}
\ea \ee
where
$$  \Mc = \ds{\lf\{\lf(\ba{ccr}n_1&\h l_1\\[1mm]
n_2&\h \l_2 \ea \ri)\ |\ n_1,n_2,l_1,l_2 \in \Z \ri\}\ .}
$$
At this point we want to stress that the set $\Mc$ is larger
than that of ref. \cite{DKL2} due to the
half--integer  entries in the right column. The restriction
on even momentum numbers in (\ref{partition}) is a direct
consequence of the definitions (\ref{complexmod}). However
it is important to note that there is no discrete transformation
on the background \cite{MS1,MS2} which allows one to return to integer moded
momentum \em and \em winding numbers.

{}From \req{syz4} one can obtain $\tau_2Z_{(\Th^2,1)}^{torus}(\tau,\bar \tau)$
by
an $S$--transformation on $\tau$. After exchanging $n_i$ and $l_i$
and performing again a Poisson resummation on $l_i$ one obtains

\be
Z_{(\Th^2,1)}^{torus}(\tau,\bar \tau,T,U)= \fc{1}{4} \sum_{m_1,m_2 \in \ZZZ
\atop n^1,n^2 \in \ZZZ}e^{2 \pi i \tau (m_1
\fc{n^1}{2}+m_2\fc{n^2}{2})} e^{\fc{-\pi \tau_2}{T_2U_2} |T U \fc{n^2}{2}+T
\fc{n^1}{2}-Um_1+m_2|^2}\ .
\label{partitionos}
\ee
The factor $1/4$ can be identified with the volume of the dual sublattice in
\req{zero}. This point has been discussed  already in \cite{ek}.
$\tau_2 Z_{(\Th^2,1)}^{torus}(\tau,\bar \tau,T,U)$ is invariant under
$\Ga^0(2)$ acting on $\tau$ and is
identical to that for the $(\Th^2,\Th^2)$ sector.
$\Ga^0(2)$ is defined similar to
\req{group} but with $b=0\ \mod\ 2$ instead of $c=0\ \mod\ 2$.
Thus the contribution of the two
sectors $(\Th^2,1)$ and $(\Th^2,\Th^2)$ to the coefficient
$b_a^{N=2}$ of the $\bet$--function is one fourth of that of the
sector $(1,\Th^2)$ and we get

\be
b^{N=2}_a=\fc{3}{2}b_a^{(1,\Th^2)}\ .
\label{bet}
\ee
The details of the integration of $Z_{(1,\Th^2)}^{torus}$
over the modular parameter $\tau$ are given in the appendix.
The final result for
the threshold correction to the inverse gauge coupling \req{thresh} reads

\bea
\triangle_a(T,\bar T,U,\bar U) = &-&\ds{b^{(1,\Th^2)}_a \ln \lf[\fc{8\pi
e^{1-\ga_E}}{3 \sqrt{3}} \fc{T_2}{4} \lf|\eta\lf(\fc{T}{2}\ri)\ri|^4 U_2
|\eta(U)|^4\ri]}\nnn
&-&\ds{\h b_a^{(1,\Th^2)}\ln \lf[\fc{8\pi
e^{1-\ga_E}}{3 \sqrt{3}} T_2 \lf|\eta\lf(\fc{T}{2}\ri)\ri|^4 U_2
|\eta(U)|^4\ri]\ .}
\label{rez4}
\eea
Formula \req{rez4} has the following interesting properties.
First the non--holomorphic moduli dependent part corresponding
to massless modes running in the loops agrees with that of
ref. \cite{DKL2} up to the modified $\bet$--function coefficient \req{bet}.
Note that $b_a^{N=2}$ depends on the torus lattice \cite{ek}.
The functional dependence of the non--holomorphic
part was expected since it determines the field theoretical anomaly
w.r.t. modular
transformations on $T$ and should be independent of the
torus lattice. The second point is the symmetry
of eq. \req{rez4} w.r.t. transformations on the moduli $T$ and $U$.
Whereas the \kae potentials \cite{clo,DKL1} are invariant under a symmetry
$\Ga_T \times
\Ga_U$ acting on $T$ and $U$, eq. \req{rez4} is only invariant under the
subgroup
$\Ga^0(2)_T \times \Ga_U$.

It is helpful to consider the contribution of the relevant two--dimensional
untwisted subspace to the energy of an
unexcited string
twisted in the remaining dimensions.
Because of twist invariance the momenta and windings w.r.t. to the other
dimensions vanish and we get

\be
m^2_\bot=\sum_{m_1,m_2 \in \ZZZ \atop n^1,n^2 \in \ZZZ} \fc{1}{T_2 U_2}|T U
n^2+T
n^1-2Um_1+2m_2|^2\ .
\label{maz4}
\ee
The common belief is that such a  formula should be  invariant under
the symmetry $\Ga_T \times \Ga_U$ acting on $T$ and $U$ respectively. But it is
not
hard to see that the above expression allows only the following transformations
on
$T$ and $U$:

\be
\ba{ccccrcr}
T &\lra& T+2 &\ , \ & T &\lra& \ds{\fc{T}{T+1}\ ,}\nnn
U &\lra& U+1 &\ , \ & U &\lra& \ds{-\fc{1}{U}\ .}
\ea
\label{duz4}
\ee
These are the generators of the group $\Ga^0(2)_T$ acting on $T$ and
$\Ga_U$ acting on $U$, respectively.
A general $\Ga$ transformation on $T$ represented by $T \ra
\fc{aT+b}{cT+d}$ requires the following
redefinition of the windings and momenta:

$$
\lf(\ba{c} n'^1\\n'^2\\m'_1\\m'_2 \ea \ri) = \lf(\ba{ccccr}
a&0&0&2c\\
0&a&-2c&0\\
0&-b/2&d&0\\
b/2&0&0&d \ea \ri)\lf(\ba{c} n^1\\n^2\\m_1\\m_2 \ea \ri)\ .$$
{}From this expression it follows immediately that $b$ has to be an even
number. Note that under the above unimodular transformation the
spin $(P_L^2-P_R^2)_\bot$ remains invariant too.
These results should be compared with the results obtained in \cite{dvv,rp} for
two--dimensional compactifications on Narain lattices.

Expression \req{maz4} can be used to consider the {\em topological free energy}
of the subspace defined by $\Lambda_\bot$ \cite{fklz}:

\be
F(T,\bar T,U,\bar U)=\sum_{m_1,m_2 \in \ZZZ \atop n^1,n^2 \in \ZZZ} \ln \fc{|T
U
n^2+T
n^1-2Um_1+2m_2|^2}{T_2 U_2}\ .
\label{free}
\ee
$F$ is the contribution of the two--dimensional unrotated
subspace to the total free energy. Since $F$ is defined to be
a topological quantity the sum of (\ref{free}) does not
run over oscillator excitations and is therefore subject to the
constraint $P_L^2-P_R^2 = 4(m_1n_1+m_2n_2) = 0$. This equation
can be solved in two inequivalent orbits \cite{ov}.
Proceeding similar as in \cite{fklz} we arrive at

\be
F(T,\bar T,U,\bar
U)=\ln\lf[\fc{T_2}{4}\lf|\eta\lf(\fc{T}{2}\ri)\ri|^4U_2|\eta(U)|^4\ri].
\label{fez4}
\ee
Interestingly
the topological free energy for $(\Lambda_\bot)^\ast$ does not agree with
\req{fez4}. It can be evaluated from \req{partitionos} to be

\be
\tilde F(T,\bar T,U,\bar U)=\h
\ln\lf[T_2\lf|\eta\lf(\fc{T}{2}\ri)\ri|^4U_2|\eta(U)|^4\ri].
\ee
Both results together agree with \req{rez4} up to the
factor $b_a^{(1,\Th^2)}$.

\subsection{Threshold corrections for the $SU(3) \times SO(8)/Z_6$ orbifold}
\label{z6}

Next we investigate
a $Z_6$ orbifold with torus lattice  $SU(3) \times SO(8)$.
The complex twist is given by $\Th=exp[\fc{2\pi i}{6}(2,1,-3)]$. Therefore
the twists $\Th^2$ and $\Th^4$ leave invariant the third complex subspace
which lies in the $SO(8)$ lattice. There is also a fixed
plane for $\Th^3$. Since this plane lies in the first two coordinates which
correspond to the $SU(3)$ lattice, the results of
\cite{DKL2} can be applied and we will consider its contribution to the
thresholds only at the end.
The twist w.r.t the lattice basis is defined to be

$$
Q=\left(\ba{ccccccr}
0&-1&0&0&0&0 \\
1&-1&0&0&0&0 \\
0&0&0&1&-1&-1 \\
0&0&1&1&-1&-1 \\
0&0&0&1&-1&0 \\
0&0&0&1&0&-1
\ea \right)\ .$$
The $Q^2$ and $Q^{\ast^2}$ invariant subplanes can be parameterized by the
following vectors, respectively:

\be
w=\lf(\ba{c} 0\\0\\n^1\\0\\n^1-n^2\\n^2
\ea \ri)\ ,\ \ n^1,n^2 \in \ZZ\ \ \ {\rm and}\ \ \ p=\lf(\ba{c}
0\\0\\m_1\\-m_1\\m_2\\m_1-m_2\ea \ri)\ ,\ \ m_1,m_2 \in \ZZ\ .
\label{vecz6}
\ee
Proceeding similar as before one obtains for the metric and
antisymmetric background fields in
the lattice basis

$$
g=\left(\ba{ccccccr}
R^2& -R^2/2& 0 &0 &0 &0\\
-R^2/2& R^2& 0& 0& 0& 0\\
0& 0& S^2 &x &z &-S^2 - 2x - z\\
0& 0& x& S^2& -S^2 - x - z& x + z\\
0& 0 &z &-S^2 - x - z& V^2& 2S^2 - V^2 + 2x + z\\
0& 0 &-S^2 - 2x - z& x + z& 2S^2 - V^2 + 2x + z&-S^2 + V^2 - 2x - 2z
\ea \right)$$
with $R,S,V,x,z \in \RR$ and
$$b=\left(\ba{ccccccr}
0&\al& 0& 0& 0& 0\\
-\al& 0& 0& 0& 0& 0\\
0& 0& 0& \bet&-\ga& \ga\\
0& 0& -\bet& 0& \bet + \ga& \bet - \ga\\
0& 0& \ga& -\bet - \ga&0&\ga\\
0& 0& -\ga& -\bet + \ga& -\ga& 0
\ea \right)\ , \ \ \ \alpha, \beta, \gamma \in \RR \ .$$
For the moduli corresponding to the third complex plane we get

\bea
U &=& \ds{\fc{1}{S^2 + V^2 + 2z}\lf[S^2 - 2V^2 - z + i \sqrt{\dete g_\bot}\ri]\
,} \nnn
T &=& \ds{2\lf[3 \ga + i \sqrt{\dete g_\bot}\ri]\ ,}
\eea
where $$\dete g_\bot:=3S^2V^2 -6S^4 - 6S^2x - 6V^2x - 12S^2z
-12xz - 9z^2\ .$$
The N=2 orbit is $\Oc=\{(1,\Th^2), (\Th^2,1), (\Th^2,\Th^2), (\Th^2,\Th^4),
(1,\Th^4), (\Th^4,1), (\Th^4,\Th^4), (\Th^4,\Th^2)\}$. The first four sectors
are the CPT
conjugate of the following ones. The fundamental orbit $\Oc_0$ consists of
$(1,\Th^2)$
and $(1,\Th^4)$. All the remaining elements of $\Oc$ can be generated by
acting on these two sectors with the group elements $S,ST$ and $ST^2$. The sum
$\tau_2Z_{(1,\Th^2)}^{torus}(\tau,\bar
\tau)+\tau_2Z_{(1,\Th^4)}^{torus}(\tau,\bar \tau)$
is invariant under the group $\Ga_0(3)$ generated by $T$ and $ST^3S$. Its
fundamental region $\tilde \Fc$ is given by the set $\tilde
\Fc=\{1,S,ST,ST^2\}\Fc$ \cite{apo}.
Acting on $\tau$ with an $ST^3S$--transformation one can show that
$Z_{(1,\Th^4)}^{torus}(\tau,\bar \tau)=Z_{(1,\Th^2)}^{torus}(\tau,\bar \tau)$
and thus
it is sufficient to integrate $2Z_{(1,\Th^2)}^{torus}(\tau,\bar
\tau)$ over the region $\tilde \Fc$.
{}From \req{zero} and \req{vecz6} we find

\be
Z_{(1,\Th^2)}^{torus}(\tau,\bar \tau,T,U)= \sum_{m_1,m_2 \in \ZZZ \atop n^1,n^2
\in \ZZZ}e^{2 \pi i \tau (m_1n_1+m_1n_2+m_2n_1-2m_2n_2)} e^{\fc{-\pi
\tau_2}{T_2U_2} |T U n^2+T
n^1-U(m_1+m_2)+m_1-2m_2|^2}\ .
\label{partitionz6}
\ee
After a Poisson transformation in the momenta and a special linear
transformation eq. \req{partitionz6} becomes

\be \label{syz6}
\ba{lrlcr}
&\tau_2 \ Z^{torus}_{(1,\Th^2)}(\tau,\bar \tau,T,U)&=& \ds{\fc{1}{3} \sum_{A
\in
\Mc} e^{- 2 \pi i T \dete A}\ T_2\
e^{\fc{-\pi T_2}{\tau_2 U'_2} \lf|(1,U')A \lf(\tau \atop 1\ri)\ri|^2}\ ,}
\ea
\ee
where
$$ \Mc = \ds{\lf\{\lf(\ba{ccr}n_1& l_1\\[1mm]
n_2&\fc{1}{3} \l_2 \ea \ri)\ |\ n_1,n_2,l_1,l_2 \in \Z \ri\}}
$$
and $U'=U+2$. Since the coefficients $b_a^{(1,\Th^2)}$ and
$b_a^{(1,\Th^4)}$ are three times larger than those
of the remaining N=2 sectors we have

\be
b_a^{N=2}=4b_a^{(1,\Th^2)}\ .
\ee
Again the details of the $\tau$--integration are given in the appendix. The
final result for
the threshold corrections \req{thresh} reads

\bea
\triangle_a(T,\bar T,U,\bar U) = &-&\ds{2b^{(1,\Th^2)}_a \ln \lf[\fc{8\pi
e^{1-\ga_E}}{3 \sqrt{3}} \fc{T_2}{3} \lf|\eta\lf(\fc{T}{3}\ri)\ri|^4
\fc{U_2}{3} \lf|\eta\lf(\fc{U+2}{3}\ri)\ri|^4\ri]}\nnn
&-&\ds{2b^{(1,\Th^2)}_a \ln \lf[\fc{8\pi
e^{1-\ga_E}}{3 \sqrt{3}} T_2 \lf|\eta\lf(T\ri)\ri|^4 U_2
\lf|\eta\lf(U+2\ri)\ri|^4\ri]}\nnn
&-&\ds{\hat{b}_a \ln \lf[\fc{8\pi
e^{1-\ga_E}}{3 \sqrt{3}} \hat{T}_2 |\eta(\hat T)|^4
\hat{U}_2 |\eta(\hat U)|^4\ri]\ .}
\label{rez6}
\eea
The third term arises from the fixed plane in the
lattice of $SU(3)$. $\hat{b}_a$ is the contribution of the sectors
$(1,\Th^3)$, $(\Th^3,1)$ and $(\Th^3,\Th^3)$ to the coefficient of the
$\beta$--function and $\hat{T}$ and $\hat{U}$ are the moduli corresponding
to this plane.

Again we consider the spectrum of strings with
non--vanishing winding
and momentum numbers only in the relevant subspace:

\be
m^2_\bot=\sum_{m_1,m_2 \in \ZZZ\atop n^1,n^2 \in \ZZZ}\fc{1}{T_2 U'_2} |T U'
n^2+T
n^1-U'm_1+3m_2|^2
\label{maz6} \ .
\ee
Its symmetry group is given by $\Ga^0(3)_T \times \Ga^0(3)_{U'}$ with
generators
\be
\ba{ccccrcr}
T &\lra& T+3 &\ , \ & T &\lra& \ds{\fc{T}{T+1}\ ,}\nnn
U' &\lra& U'+3 &\ , \ & U' &\lra& \ds{\fc{U'}{U'+1}\ .}
\ea
\label{duz6}
\ee
The above transformations accompanied by a proper unimodular transformation on
the winding and momentum numbers keep invariant the spin $(P_L^2-P_R^2)_\bot$.
The free energy of $\Lambda_N^\bot$ can be calculated to be

\be
F(T,\bar T,U,\bar
U)=\ln\lf[\fc{T_2}{3}\lf|\eta\lf(\fc{T}{3}\ri)\ri|^4U_2\lf|\eta\lf(U
+2\ri)\ri|^4\ri]+\ln\lf[T_2\lf|\eta(T)\ri|^4\fc{U_2}{3}\lf|\eta\lf(\fc{U
+2}{3}\ri)\ri|^4\ri] \ ,
\label{fez6}
\ee
and similarly for $(\Lambda_N^\bot)^{\ast}$:

\be
\tilde F(T,\bar T,U,\bar
U)=\ln\lf[T_2\lf|\eta\lf(T\ri)\ri|^4U_2|\eta(U+2)|^4\ri]+\ln\lf[\fc{T_2}{3}\lf|\eta\lf(\fc{T}{3}\ri)\ri|^4\fc{U_2}{3}\lf|\eta\lf(\fc{U
+2}{3}\ri)\ri|^4\ri]
\ee
Both results together agree with \req{rez6} up to the factor
$b^{(1,\Th^2)}_a$.

\subsection{Threshold corrections for the $SU(2) \times SO(10)/Z_8$ orbifold}
\label{z8}

The $Z_8$--IIa orbifold with torus lattice $SU(2) \times SO(10)$
has the complex twist $\Th=exp[\fc{2\pi i}{8}(1,3,-4)]$. W.r.t the lattice
basis
vectors
the twist reads

\be
Q=\left(\ba{ccccccr}
0&0&1& -1& -1& 0\\
1& 0&1&-1& -1& 0\\
0& 1& 1& -1& -1& 0\\
0& 0& 1& -1& 0& 0\\
0& 0& 1& 0& -1& 0\\
0& 0& 0& 0& 0& -1
\ea \right)\ .
\ee
For the metric $g$ and antisymmetric background field $b$ we
find

$$
g=\left(\ba{ccccccr}
R^2& x& 0& -R^2/2 - x& -R^2/2 - x& 0\\
x& R^2& x& -R^2/2 - x& -R^2/2 - x& 0\\
0& x& R^2& -R^2/2& -R^2/2& 0\\
-R^2/2 - x& -R^2/2 - x& -R^2/2&S^2&2R^2 - S^2 + 2x& -u\\
-R^2/2 - x& -R^2/2 - x& -R^2/2& 2R^2 - S^2 + 2x& S^2&u\\
0& 0& 0& -u&u& V^2
\ea \right)\ ,$$

$$
b=\left(\ba{ccccccr}
0& \al& -2\al + 2\bet & \al - \bet& \al - \bet&0\\
-\al& 0& \al& -\al + \bet& -\al + \bet& 0\\
2\al - 2\bet& -\al& 0& \bet&\bet& 0\\
-\al + \bet&\al - \bet&-\bet& 0& 0& -\ga\\
-\al + \bet& \al - \bet& -\bet& 0&0&\ga\\
0& 0& 0& \ga& -\ga& 0
\ea \right)\ $$
with $R,S,V,x,u,\alpha,\beta,\gamma \in \RR$.
$Q^2,\ Q^4$ and $Q^6$ leave invariant the third complex subspace. This subspace
is spanned by the following basis
vectors:

\be
w=\lf(\ba{c} 0\\0\\0\\n^1\\-n^1\\n^2
\ea \ri)\ ,\ \ n^1,n^2 \in \ZZ\ \ \ {\rm and}\ \ \ p=\lf(\ba{c}
0\\0\\0\\m_1\\-m_1\\m_2\ea \ri)\ ,\ \ m_1,m_2 \in \ZZ\ .
\label{vecz8}
\ee
The $(1,1)$-- and $(2,1)$--modulus belonging to deformations of the third
complex coordinates can be defined as

\bea
T &=& 4\lf[-\ga +i\sqrt{-u^2 + V^2(S^2 -R^2- x)}\ri]\ ,\nnn
U &=& \ds{\fc{1}{2S^2-2R^2 - 2x}\lf[-u +i\sqrt{-u^2 + V^2(S^2 -R^2- x)}\ri]\ .}
\eea
Using \req{vecz8} we get for $Z_{(1,\Th^2)}^{torus}$ in \req{zero} the
following
expression:

\be
Z_{(1,\Th^2)}^{torus}(\tau,\bar \tau,T,U)= \sum_{{m_1, m_2 \in \ZZZ}\atop
n^1,n^2 \in \ZZZ}e^{2
\pi i \tau (2 m_1
n^1+m_2n^2)} e^{\fc{-\pi \tau_2}{T_2 U_2} |T U n^2+T
n^1-2Um_1+m_2|^2}\ .
\label{partitionz8}
\ee
After a Poisson transformation in the momenta \req{partitionz8} becomes

\be \label{syz8}
\ba{lrlcr}
&\tau_2 \ Z_{(1,\Th^2)}^{torus}(\tau,\bar \tau,T,U)&=& \ds{\fc{1}{2} \sum_{A
\in
\Mc} e^{- 2 \pi i T \dete A}\ T_2\
e^{\fc{-\pi T_2}{\tau_2 U_2} \lf|(1,U)A \lf(\tau \atop 1\ri)\ri|^2}\ ,}
\ea
\ee
where
$$\Mc = \ds{\lf\{\lf(\ba{ccr}n_1& \h l_1\\[1mm]
n_2& \l_2 \ea \ri)\ |\ n_1,n_2,l_1,l_2 \in \Z \ri\}\ .}
$$
The N=2 orbit is
$\Oc=\{(1,\Th^2),(\Th^2,1),(\Th^2,\Th^2),(\Th^2,\Th^4),(1,\Th^6),(\Th^6,1),(\Th^
6,\Th^6),(\Th^6,\Th^4),\\
(\Th^4,\Th^6),(\Th^6,\Th^2),
(\Th^4,\Th^2),(\Th^2,\Th^6)\} \cup
\{(1,\Th^4),(\Th^4,1),(\Th^4,\Th^4)\}\ .$ The second set is
invariant under
$\Ga$ by itself. Moreover it is a irreducible representation of $\Ga$. The
fundamental orbit consists of
$\Oc_0=\{(1,\Th^2),(1,\Th^4),(1,\Th^6),(\Th^4,\Th^2),(\Th^4,\Th^6)\}$. All
other sectors can be generated from these by acting on $\tau$ with $1,S$ and
$ST$.
The partition functions for all sectors appearing in $\Oc_0$ agree
with the one of $(1,\Th^2)$. The coefficient $b_a^{(1,\Th^2)}$
 of the untwisted sector
is two times that of the twisted sectors $(\Th^2,1)$ and $(\Th^2,\Th^2)$.
We postpone the further details to the appendix and find
the following expression for the threshold corrections
\req{thresh}:

\bea
\triangle_a(T,\bar T,U,\bar U)= &-& \ds{5b_a^{(1,\Th^2)} \ln \lf[\fc{8\pi
e^{1-\ga_E}}{3 \sqrt{3}} T_2 \lf|\eta\lf(T\ri)\ri|^4 U_2 |\eta(2U)|^4\ri]}\nnn
&-&\ds{5b_a^{(1,\Th^2)} \ln \lf[\fc{8\pi
e^{1-\ga_E}}{3 \sqrt{3}} T_2 \lf|\eta\lf(\fc{T}{2}\ri)\ri|^4 U_2
|\eta(U)|^4\ri]\ .}
\label{rez8}
\eea
Again this result is no longer invariant under a symmetry
$\Ga_T \times \Ga_U$ but under the group $\Ga^0(2)_T \times \Ga_0(2)_U$. This
symmetry group agrees with the one of the
spectrum: The mass formula for the two--dimensional subspace
reads

\be
m^2_\bot=\sum_{m_1,m_2 \in \ZZZ\atop n^1,n^2 \in
\ZZZ}\fc{1}{T_2 U_2} |T U n^2+T
n^1-2Um_1+m_2|^2\ .
\label{maz8}
\ee
This formula is invariant under the following transformations
on $T$ and $U$, respectively:

\be
\ba{ccccrcr}
T &\lra& T+2 &\ , \ & T &\lra& \ds{\fc{T}{T+1}\ ,}\nnn
U &\lra& U+1 &\ , \ & U &\lra& \ds{-\fc{U}{2U-1}\ .}
\ea
\label{duz8}
\ee
Together they form the generators of the group $\Ga^0(2)_T \times \Ga_0(2)_U$.
The topological free energy can be calculated to be

\bea
F(T,\bar T,U,\bar
U)&=&\ds{\ln\lf[T_2\lf|\eta\lf(T\ri)\ri|^4U_2\lf|\eta\lf(2U\ri)\ri|^4\ri]+\ln\lf
[\fc{T_2}{2}\lf|\eta\lf(\fc{T}{2}\ri)\ri|^4U_2|\eta(U)|^4\ri]\
,}\nnn
\tilde F(T,\bar T,U,\bar
U)&=&\ds{\ln\lf[T_2\lf|\eta\lf(\fc{T}{2}\ri)\ri|^4U_2|\eta(2U)|^4\ri]+\ln\lf[T_2
\lf|\eta\lf(T\ri)\ri|^4U_2\lf|\eta\lf(U\ri)\ri|^4\ri]\ ,}
\eea
for the two inequivalent lattices $\Lambda_{N}^\bot$ and
$(\Lambda_{N}^\bot)^\ast$ discussed previously, respectively.

\sect{Anomaly cancellation and gauge coupling unification}

In the effective field theory
the non--holomorphic part of the threshold corrections calculated before (see
\req{rez4}, \req{rez6} and \req{rez8}) is reproduced by triangle graphs
involving two gauge fields and a connection of the \kae or
sigma--model coordinate transformations
\cite{dfkz,l,co}.
The contribution of these graphs can be easily calculated
by considering its supersymmetric completion, the parity odd
coupling representing mixed gauge--\kae or gauge--sigma model coordinate
anomalies. In the presence of a Green--Schwarz mechanism, that is, a one-loop
coupling of the antisymmetric tensor field to some gauge or auxiliary gauge
field, one has to substract its contribution from the total anomaly
coefficient. For a generic modulus the one-loop correction to
the gauge coupling is then given by

\be
\triangle_{a,FT} = (c_{\rm K}^a + c_{\rm C}^{A,a} - \delta_{\rm GS})\
\ln (\im \ T) \ .
\label{simhoibesanahaschnizl}
\ee
Here $c_{\rm K}^a$ and $c_{\rm C}^{A,a}$ are group--theoretical
coefficients characterizing the contribution of the massless spectrum
due to the coupling to the \kae and sigma--model connection,
respectively and $\delta_{\rm GS}$ is the coefficient in front
of the gauge group independent Green--Schwarz counterterm. For further details
we refer the reader to refs. \cite{dfkz,l,co}. $\delta_{\rm GS}$
can be determined from the previous string calculation to be
\be
\delta_{\rm GS} = c_{\rm K}^a + c_{\rm C}^{A,a} - b_a^{N=2} \ .
\ee
Obviously $\triangle_{a,FT}$ is not invariant under modular transformations
on the modulus $T$, but only the full expression for the threshold correction
including the holomorphic part. Since we have shown in the previous
section that the symmetry group of the threshold corrections w.r.t.
reparametrizations of the moduli manifold is only a subgroup of the full
modular group $\Ga$, it is clear that the exact expression could not be
inferred from
the field--theoretical term \req{simhoibesanahaschnizl} by imposing
anomaly cancellation. This is
because the relevant subgroup has no longer a single automorphic
function as it happened in the case of the group $\Ga$. Therefore the
previous string calculations are necessary to determine the
exact form of the full threshold corrections which provide the
boundary conditions for the running coupling constants.
The renormalization group equation is given by \cite{kap,dfkz,ILR}

\be \label{run}
\fc{1}{g^2_a(\mu)}=\fc{k_a}{g^2_{\rm string}}+\fc{b_a}{16 \pi^2} \ln \fc{M_{\rm
string}^2}{\mu^2}
-\fc{1}{16\pi^2}\triangle_a \ ,
\ee
where $k_a$ is the Kac--Moody level of the gauge group $a$
and $b_a$ is the complete $\bet$--function coefficient of
all orbifold sectors. $M_{\rm string}$ denotes the string scale of the order
of the Planck scale. In ref. \cite{kap} it was calculated to be
$M_{\rm string} = 0.7 \, g_{\rm string} \times 10^{18}\, {\rm GeV}$ in the
$\overline{\rm DR}$ scheme. $\triangle_a$ is given by \req{rez4},\req{rez6} and
\req{rez8} for the $Z_4$, $Z_6$ and $Z_8$ orbifold models considered in
the previous sections, respectively.
The unification mass, defined as the scale where the
two gauge couplings $g_a$ and $g_b$ meet
can be determined to be
\be
\begin{array}{ll}
Z_4 : \,
&M_X=\ds{M_{\rm string}
\lf[\fc{U_2T_2}{4^{2/3}}\lf|\eta\lf(\fc{T}{2}\ri)\ri|^4|\eta(U)|^4\ri]^{\fc{b^{N
=2}_b-b^{N=2}_a}{2(b_a-b_b)}} \ ,} \nnn
\\
Z_6 : \, &M_X =\ds{M_{\rm
string}\lf[\fc{T_2U_2}{3}\lf|\eta(T)\eta\lf(\fc{T}{3}\ri)\eta(U+2)\eta\lf(\fc{U+
2}{3}\ri)\ri|^2\ri]^{\fc{b^{N
=2}_b-b^{N=2}_a}{2(b_a-b_b)}} \ ,}
\nnn \\
Z_8 : \, &M_X =\ds{M_{\rm
string}\lf[T_2U_2\lf|\eta(T)\eta\lf(\fc{T}{2}\ri)\eta(U)\eta\lf(2U\ri)\ri|^2\ri]^{\fc{b^{N
=2}_b-b^{N=2}_a}{2(b_a-b_b)}} \ .} \nnn
\end{array}
\ee
The new functional dependence of the threshold corrections has to be
taken into account in phenomenological discussions of the above
orbifold models. In refs. \cite{ILR,IL} numerous orbifolds have been
investigated
with respect to their predictions for the values of the weak mixing angle
$\sin^2 \th_W$ and the strong coupling constant
$\al_{\rm s}$ assuming  the particle content of a
minimal supersymmetric standard model.
Moreover the various models have been classified
using the constraints on the spectrum following from
anomaly freedom with respect to duality transformations on the
moduli in the planes rotated by all orbifold twists.
It is remarkable that
although the explicit form of the threshold corrections evaluated above
differs from the expressions for the known orbifolds,
most of the arguments of ref. \cite{IL} go through.
Since the symmetry group of the
moduli of the unrotated plane is a subgroup of the modular group,
the modular
weights for the matter fields do not change. Indeed the conditions for the
modular weights arising from requiring the
correct weak mixing angle and strong
coupling constant together with
anomaly freedom with respect to the moduli reparametrizations
in the completely rotated planes remain valid. What will change
are the values for the moduli $T$ and $U$ which solve the minimal
unification scenario. The fact that $T$ and $U$ are divided
by integer numbers in the threshold corrections favours the tendency
to higher values and raises the problem of solutions
far away from the self--dual point.

\vskip 1cm
\centerline{\bf Summary and Conclusions}
\ \\
We derived the threshold corrections to the gauge couplings for
the class of toroidal orbifolds models with a generic six--dimensional
torus lattice which does not
split into a direct sum of orthogonal sublattices.
The final expressions (cf. \req{rez4}, \req{rez6} and \req{rez8}) do not have
the functional form which has been expected by
just considering the symmetry of the \kae potentials:
The thresholds
are not invariant under the group $\Ga_T \times \Ga_U$ acting on the
relevant moduli $T$ and $U$. In contrast they turn out to be
functions automorphic only under a subgroup of $\Ga_T \times \Ga_U$. In
addition we showed that the spectrum shares this property. This has
interesting consequences for the Yukawa couplings between twisted matter
fields:
the coupling w.r.t. the unrotated plane can be shown to reduce always to
a twist--antitwist coupling as a consequence
of the point group selection rule \cite{ejss}. Since its
functional dependence on the moduli is the same as that of the spectrum,
this special class of
Yukawa couplings will have exactly the same symmetry properties
as the spectrum and the threshold corrections.

While the non--holomorphic part of the threshold corrections is fixed
by the known K\"ahler potentials up to constant terms, the holomorphic
part which describes the
massive string states and is necessary for the
cancellation of duality anomalies
in the effective theory has
changed compared to the results of ref. \cite{DKL2,dfkz}. In contrast
formulas where only the
weights of modular functions appear, like the transformation behaviour
of the vertex operators of the various fields or the superpotential
\cite{flst},
remain untouched by the change of the symmetry group.
We want to mention that the symmetry of the spectrum of
the completely untwisted
sector is not determined by the above considerations. In general
a mass formula corresponding to e.g. \req{maz4} can have quite a complicated
dependence on the whole set of moduli fields. This was shown in
ref. \cite{es} for the case of the $Z_7$ orbifold.

Our above calculations indicate that the significance of the threshold
corrections is far from beeing investigated sufficiently. Just a different
choice of the lattice already has changed the results in a remarkable way.
In more general ---and more realistic--- models, like orbifolds with
Wilson lines \cite{hpn} or vacuum exspectation values of certain scalar
fields, there will arise additional contributions to the thresholds of
an up to now unknown size. This fact might change the phenomenological
prospects of such models.

\vskip 1cm
\centerline{\bf Acknowledgements}
\vskip 0.5cm

We would like to thank Lance Dixon, Albrecht Klemm and especially Hans Peter
Nilles for helpful
discussions.

\vskip 1.5cm

\section*{Appendix}

\appendix
\renewcommand{\theequation}{\arabic{equation}}
\setcounter{equation}{0}

In this appendix we present the technical details necessary
for evaluating the
integral and infinite sum in the previous formulas.

\section*{$Z_4$ orbifold}

Under a $\Ga_0(2)$ transformation $\tau \ra \fc{a\tau+b}{c\tau+d}$
(with $ad-bc=1\ ,\ c=0\ \mod\ 2$), eq. \req{syz4} remains invariant if we
transform
the four integral numbers $n_1,n_2,l_1$ and $l_2$
in the following way:

\be
\lf(\ba{cc} n'_1& n'_2 \\
                l'_1& l'_2 \ea \ri)=\lf(\ba{cc} a& c/2 \\
                                              2b&d \ea \ri)
\lf(\ba{cc} n_1& n_2 \\
              l_1& l_2 \ea \ri) \ .
\label{trans}
\ee
Clearly \req{syz4} is only
invariant under the group $\Ga_0(2)$ acting on $\tau$.
To proceed similar as in the appendix of ref. \cite{DKL2} we have to
divide the
set of all integral two by two matrices into equivalence classes under the
group $\Ga_0(2)$. There are three kinds of orbits:
\begin{enumerate}
\item The zero matrix whose contribution to \req{syz4} and the integral
\req{thresh} is denoted by $I_1$.
The integration over $\tilde \Fc=\{1,S,ST\}\Fc$ yields

$$I_1=\fc{3}{2} \times \fc{\pi}{3} \fc{T_2}{2}\ .$$
\item All matrices with non--zero determinants. Since
the fundamental region of $\Ga_0(2)$ is $\tilde \Fc=\{1,S,ST\}\Fc$ we obtain
the
following representatives:

\be
\lf(\ba{cc} k& j \\ 0& p \ea \ri)\ ,\ \lf(\ba{cc} 0& -p \\ k& j \ea
\ri)\ ,\ \lf(\ba{cc} 0& -p \\ k& j+p \ea \ri)\ ,\ \ 0\leq j < k
, \ \ p \neq 0\ .
\label{i2m}
\ee
One can generate all
$GL(2,\ZZ)$ matrices from these three matrices by acting\footnote{To get also
the matrices with an overall minus
sign one has to
make the change $(a,b,c,d) \ra (-a,-b,-c,-d)$ which gives an
additional factor of two (cf. \cite{DKL2}).}
on them with $\Ga_0(2)$.
Their contribution to \req{thresh} together with that of $I_1$
is

$$I_1+I_2=-\fc{3}{2} \times 4 \re \ln \eta\lf(\fc{T}{2}\ri)\ .$$

\item The orbits of matrices with zero determinant can be obtained from the two
representatives

\be
\lf(\ba{cc} 0& 0 \\ j& p \ea \ri)\ ,\ \lf(\ba{cc} j& p \\ 0& 0 \ea
\ri)\ ,\ j,p \in \ZZ\ ,\ (j,p) \neq (0,0)\ .
\label{i3m}
\ee
Note that not all $\Ga_0(2)$ group elements acting on \req{i3m} lead to
different matrices. A $T^n$--transformation on $\tau$
does not modify the first of the above matrices because this
change in $\tau$ is accompanied by a multiplication with $\lf(\ba{cc} 1& 0 \\
2n& 1 \ea \ri)$ on the matrix.
Similar a $ST^{2n}S$ transformation on $\tau$ does not change the second
representative. Following \cite{DKL2} the contribution $I_3$ of the third class
together with
the last term of \req{thresh} can be calculated to be

\bdm
\ba{ll}
I_3=&\ds{-4 \re \ln \eta(U)-\ln
\lf(\fc{T_2}{4}U_2\ri)+\lf(\ga_E-1-\ln \fc{8\pi}{3\sqrt{3}}\ri)}\\[4mm]
&\ds{-\h \times 4 \re \ln \eta(U)-\h \times \ln
\lf(T_2U_2\ri)+\h \times \lf(\ga_E-1-\ln \fc{8\pi}{3\sqrt{3}}\ri)\ .}
\ea
\edm
The first matrix in \req{i3m} has to be
integrated over the half--band $\{\tau \in \CC\ |\ \tau_2>0\ ,\ |\tau_1|<\h\}$
as explained in ref. \cite{DKL2}. In contrast
the second matrix has to be integrated over a
half--band with the double width in $\tau_1$.

\end{enumerate}

\section*{$Z_6$ orbifold}

Acting with $\Ga_0(3)$ on $\tau$ leaves invariant eq. \req{syz6}
if
the momentum and winding numbers are transformed unimodularly as

\bdm
\ba{lcrclcr}
n_1' &=& \ds{an_1+cl_1} &\ ,\ & n_2'&=&\ds{an_2+\fc{c}{3}l_2 \ ,}\\[4mm]
l_1' &=&bn_1+dl_1 &\ ,\ &l_2' &=& 3bn_2+dl_2\ .
\ea
\edm
Again these transformations can be written in the form of \req{trans}
restricting to $l_1 \in 3 \ZZ$. Moreover the
exponential in \req{syz6} has to be replaced by

$$\ds{e^{-2 \pi i T(n_1\fc{l_2}{3}-n_2\fc{l_1}{3})} e^{\fc{-\pi T_2}{\tau_2
U_2'}|n_1\tau+\fc{l_1}{3}+U'n_2\tau+U'\fc{l_2}{3}|^2}\ ,\ l_1 \in 3\ZZ\ .}$$
For $\Ga_0(3)$ there is a fourth orbit additional to
those of \req{i2m}.
It is generated by the matrix

$$\lf(\ba{cc} 0& -p \\ k& j+2p \ea \ri)\ \ ,\ \ 0\leq j <k\ \ ,\ \ p \neq 0\
.$$
The four matrices with non--zero determinants together with the
zero matrix give
rise to the contribution

$$I_1+I_2=-4 \re \ln \eta\lf(\fc{T}{3}\ri)-4 \re \ln
\eta(T)\ .$$
\ \\
Since the second matrix of \req{i3m} remains invariant under a
$ST^{3n}S$--transformation on $\tau$, the contribution of this matrix in
\req{thresh} has to be integrated over three neighbouring
half--bands.
In this way one obtains for the matrices with zero--determinants the following
result:

\bdm
\ba{lcr}
I_3=&\ds{-4 \re \ln \eta\lf(\fc{U'}{3}\ri)-\ln
\lf(\fc{T_2}{3}\fc{U_2}{3}\ri)+\lf(\ga_E-1-\ln \fc{8\pi}{3\sqrt{3}}\ri)}\\[4mm]
&\ds{-4 \re \ln \eta(U')-\ln
\lf(T_2U_2\ri)+\lf(\ga_E-1-\ln \fc{8\pi}{3\sqrt{3}}\ri)\ .}
\ea
\edm

\section*{$Z_8$ orbifold}

A $\Ga_0(2)$ transformation on $\tau$ is equivalent to the following
unimodular redefinition of the integral numbers in \req{syz8}:

\bdm
\ba{lcrclcr}
n_1' &=& \ds{an_1+\fc{c}{2}l_1} &\ ,\ & n_2'&=&an_2+cl_2
\ ,\\[4mm]
l_1' &=&2bn_1+dl_1 &\ ,\ &l_2' &=& bn_2+dl_2\ .
\ea
\edm
Again this expression can be written similar as in
\req{trans} if one restricts $l_2$ to even
numbers. The exponentials of \req{syz8} have to be modified
to

$$\ds{e^{-2 \pi i T(n_1\fc{l_2}{2}-n_2\fc{l_1}{2})} e^{\fc{-\pi T_2}{\tau_2
U_2}|n_1\tau+\fc{l_1}{2}+Un_2\tau+U\fc{l_2}{2}|^2}\ ,\ l_2 \in 2\ZZ\ .}$$
For the orbits we can take \req{i2m} and \req{i3m} but with even $l_2$.
Their contribution together with that of the zero matrix can be
calculated to be

$$I_1+I_2=-2 \times 4 \re \ln \eta\lf(\fc{T}{2}\ri)-\h \times 4 \re \ln
\eta(T)\ .$$
The integral from the two orbits with zero determinant is:

\bdm
\ba{lcr}
I_3=&\ds{-4 \re \ln \eta(2U)-\ln
\lf(T_2U_2\ri)+\lf(\ga_E-1-\ln \fc{8\pi}{3\sqrt{3}}\ri)}\\[4mm]
&\ds{-4 \re \ln \eta(U)-\ln
\lf(T_2U_2\ri)+\lf(\ga_E-1-\ln \fc{8\pi}{3\sqrt{3}}\ri)}\ .
\ea
\edm

\end{document}